\documentclass[graybox]{svmult}

\usepackage{type1cm}        
\usepackage{makeidx}         
\usepackage{graphicx}        
\usepackage{multicol}        
\usepackage[bottom]{footmisc}

\usepackage{newtxtext}       %
\usepackage{newtxmath}       

\makeindex             

\begin{document}

\title*{Multi-Agent Learning for Resilient Distributed Control Systems}
\author{Yuhan Zhao \and Craig Rieger \and Quanyan Zhu}
\authorrunning{Y. Zhao \and C. Rieger \and Q. Zhu}
\institute{Yuhan Zhao \at New York University, Brooklyn, NY 11201, USA \email{yhzhao@nyu.edu}
\and Craig Rieger \at Idaho National Laboratory, Idaho Falls, ID 83402, USA \email{name@email.address}
\and Quanyan Zhu \at New York University, Brooklyn, NY 11201, USA \email{qz494@nyu.edu}}
%
%
\maketitle

\abstract{Resilience describes a system's ability to function under disturbances and threats. Many critical infrastructures, including smart grids and transportation networks, are large-scale complex systems consisting of many interdependent subsystems. Decentralized architecture becomes a key resilience design paradigm for large-scale systems. In this book chapter, we present a multi-agent system (MAS) framework for distributed large-scale control systems and discuss the role of MAS learning in resiliency. This chapter introduces the creation of an artificial intelligence (AI) stack in the MAS to provide computational intelligence for subsystems to detect, respond, and recover. We discuss the application of learning methods at the cyber and physical layers of the system. The discussions focus on distributed learning algorithms for subsystems to respond to each other, and game-theoretic learning for them to respond to disturbances and adversarial behaviors. The book chapter presents a case study of distributed renewable energy systems to elaborate on the MAS architecture and its interface with the AI stack.}

\section{Introduction} \label{sec:introduction}
Industrial Control Systems (ICSs) are central to the control and command of critical infrastructure systems, such as power grids, transportation networks, heavy industries, and distribution systems. An ICS consists of interconnected subsystems of different functions, including supervisory control and data acquisition (SCADA) systems for data collection and monitoring; field controller systems for physical plant control. Based on the unit functionality, each subsystem contains Informational Technology (IT) units (constituting IT systems) for information processing, such as task scheduling, and Operational Technology (OT) units (constituting OT systems) for real-time task execution. The subsystems work in a distributed but interdependent fashion to achieve the overall functionality of the ICS and support critical infrastructure services.

However, current ICS environments still contain many legacy systems. The organic incorporation of different control system vendors and models over the years exposes a large attack surface  to the attacker, making ICSs more vulnerable to malicious attacks. With cyber attacks becoming more technologically advanced, many of the systems have become the foremost target of resourceful attacks \cite{rieger2019industrial,zhu2021cybersecurity}. Cyber risks can also propagate among the subsystems due to extensive interconnection and interdependencies, which are not always known or planned for. Unexpected errors can happen in one subsystem when other subsystems are attacked. It is challenging to secure ICS holistically at a system-wide level. In addition to the external attack threats, the internal faults can also propagate over the ICS network, causing cascading errors and exacerbating failures in the system \cite{zhu2012dynamic,hayel2015resilient}.

The past few years have witnessed many calamitous attacks on ICSs ranging from the Stuxnet attack on nuclear power plants in 2010 \cite{farwell2011stuxnet} to the ransomware attack on Automotive technology manufacturer, Denso, in 2022 \cite{denso2022cpo}. There is an urgent need to strengthen the security of ICSs. Mere IT solutions alone are, however, insufficient to address the security challenges because they overlook the security of the OT systems. The OT systems control physical units such as sensors and actuators for task execution, and their performance is affected by the well-being of cyber systems as well as the disturbances and internal faults on physical units. The OT security aims to not only mitigate the malicious attacks from the cyber domain but also stabilize the physical systems. Therefore, a holistic cyber-physical or joint IT-OT security solution is in need to address the ICS security challenges.

ICS security is challenging, not only due to the cross-layer and system-wide needs but also because of the role of the attacks. Growingly sophisticated attacks make it harder or unfeasible to achieve ``perfect" security, which aims to protect ICSs from any attacks and ensure system performance. Resilience becomes an effective way to further mitigate the security risk. Resilience refers to the system's ability to ``maintains state awareness and an accepted level of operational normalcy in response to disturbances, including threats of an unexpected and malicious nature"  \cite{rieger2009resilient}. A resilient ICS can effectively respond to an unknown attack to mitigate its impact and maintain acceptable system performance \cite{zhu2011hierarchical,rieger2013hierarchical,zhu2020cross}. 
The resilience of ICSs needs to be achieved in a distributed way since ICSs are large-scale and distributed systems that consist of different subsystems. Each functional subsystem requires resilience to maintain its function under external attacks and internal disturbances, which consequently contributes to the resilience of the holistic ICS and the mission criticality. One plausible architecture to achieve it is through the concept of multi-agent resilience (MAR). 

\begin{figure}
    \centering
    \includegraphics[scale=0.35]{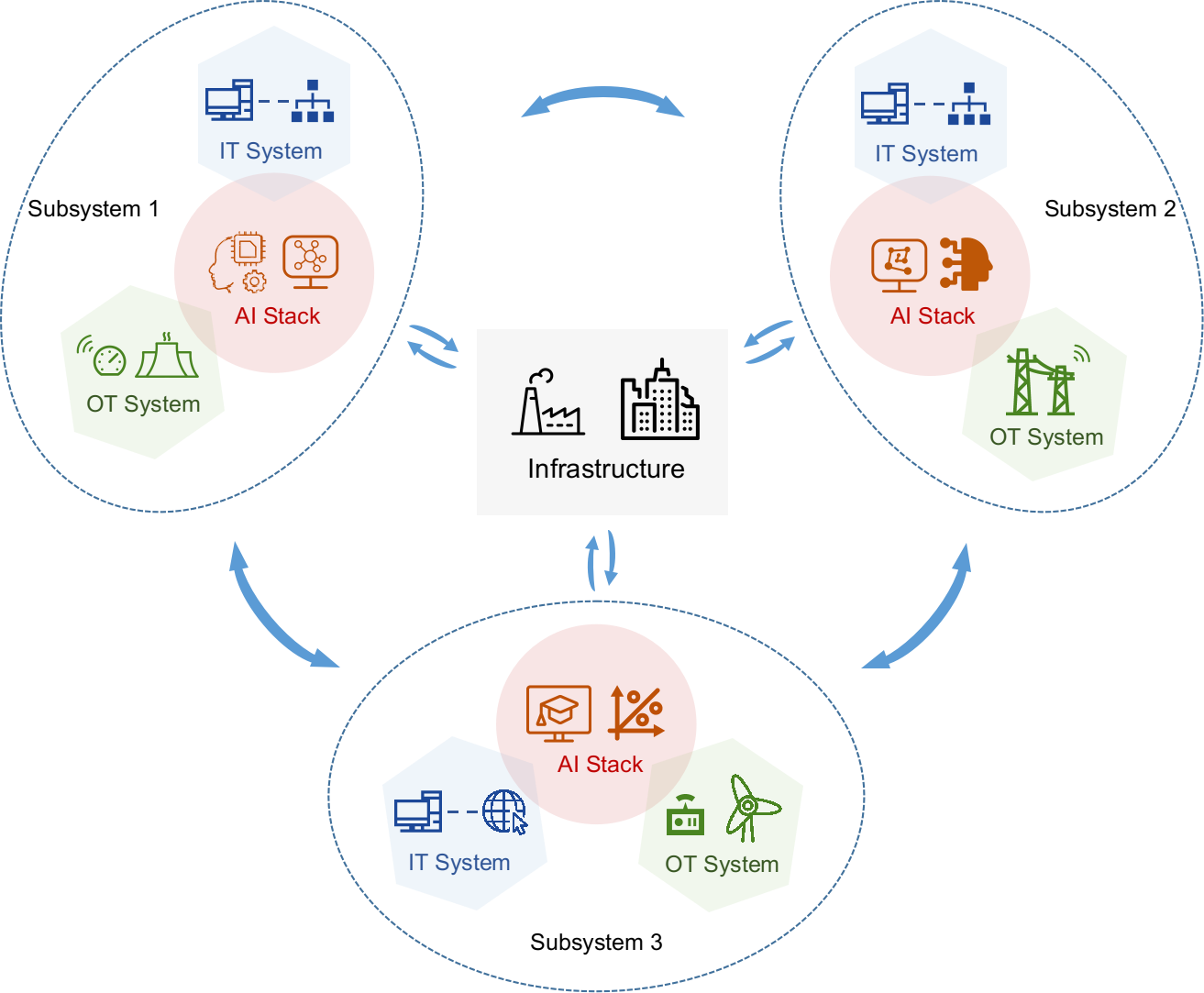}
    \caption{The illustration of an AI-augmented MAS framework. A modern ICS consists of multiple cyber-physical subsystems acting as intelligent agents. At the system level, the subsystems interact and cooperate to accomplish system-wide missions. At the subsystem level, the IT and OT systems behave as intelligent agents who manage and control the tasks in the cyber and physical domains.
    The AI stack in each subsystem interacts with not only the IT and OT systems within the subsystem but also other functional subsystems, which enables data analytics for decision-making and provides solutions to support critical infrastructures.}
    \label{fig:intro}
\end{figure}

MAR builds upon the multi-agent system (MAS) framework for large-scale control systems shown in Fig.~\ref{fig:intro}, which describes ICSs at different system scales. At the system level, each functional subsystem is treated as an intelligent agent. They interact and cooperate with each other to accomplish a system-wide mission. At the subsystem level, the IT and OT systems in each subsystem act as intelligent agents to manage, control, and coordinate the tasks in the cyber and physical domains.
It is important to note that MAR is not a mere replica of one resilient subsystem because of the heterogeneity and interdependency in ICSs \cite{rieger2012agent}. First, subsystems are diverse in their functionality and network structures, resulting in different strategies and designs for subsystems to achieve resilience. Second, the extensive interconnections among subsystems in ICSs lead to complex interdependencies. A response of one subsystem to mitigate attack consequences may exacerbate the performance of another. Therefore, MAR needs a careful and holistic set of design methodologies \cite{zhu2020cross}.

It is challenging to achieve MAR in ICSs because subsystem models and interdependencies between subsystems are not always known or planned for. Distributed learning-based mechanisms provide a possible and promising way to reach it.
First, the distributed learning-based mechanism uses data analytics for informed decision-making, which is prominent when the exact model and interdependency are increasingly challenging to obtain with the growing of ICS scale and complexity. The data for analysis is readily supported and collected by SCADA systems.
Second, the learning-based mechanism enables a continuous cognitive capability and creates an adaptive response to threats \cite{huang2022reinforcement}. A subsystem can constantly update its decision-making model to adapt to evolving changes from attackers through learning.
Third, the learning-based mechanism allows responding to unanticipated events. The learning enables each subsystem to detect and recognize new anomalies and faults in time during the operation, which is critical for protecting the system from zero-day attacks.
Finally, the learning-based mechanism allows sharing of intelligence among or within subsystems to improve the holistic ICS resilience. Instead of solely focusing on a single subsystem resilient design, the distributed learning also enables the perception and adaption of one subsystem to others. With the exchange of local knowledge \cite{zhu2009game,fung2010bayesian,zhu2012guidex,fung2011smurfen}, the subsystems are able to handle the complex and unknown interdependencies and avoid the counter-effect of resilient response interplay between subsystems.

In this chapter, we introduce a multi-agent learning framework to achieve resilient design for large-scale and distributed ICSs. Compared with traditional ICSs, an AI stack is introduced to create a learning-based resilience mechanism, which enables data-driven decision-making and adaption to unanticipated attacks and system faults for all subsystems in ICSs. 
The multi-agent learning mechanism can be designed at different system scales based on our MAS framework illustrated in Fig.~\ref{fig:intro}. At the system level, the subsystems of an ICS act as intelligent agents and learn in a distributed way to achieve system-wide resilience. Each subsystem learns to respond to threats resiliently within its sphere of influence and share intelligence through the distributed learning paradigm. At the subsystem level, the IT and OT systems in each subsystem also use learning to mitigate attacks and disturbances within the subsystem and create a cyber-physical co-learning paradigm due to the natural interdependency between IT and OT systems. This cyber-physical co-learning coupled with the distributed learning paradigm constitutes the framework of multi-agent learning mechanisms for resilient ICSs.

The chapter is organized as follows. Section \ref{sec:ics} introduces the concepts and MAS framework of ICSs and summarizes the recent emerging security challenges in ICSs. Section \ref{sec:resilience} discusses the distributed structure of resilience design in ICSs and introduces an AI stack for creating a learning-based resilience mechanism. In Section \ref{sec:learning}, we first propose the feedback learning structure in the IT and OT systems of ICSs. Then, we provide an overview of the existing learning-based literature. We elaborate  further on the learning-based resilience framework using a case study of distributed energy systems in modern power grids in Section \ref{sec:case}. Section \ref{sec:conclusion} concludes the chapter.

\section{Security Challenges of Industrial Control Systems} \label{sec:ics}
\subsection{Industrial Control Systems}
\begin{figure}
    \centering
    \includegraphics[scale=0.3]{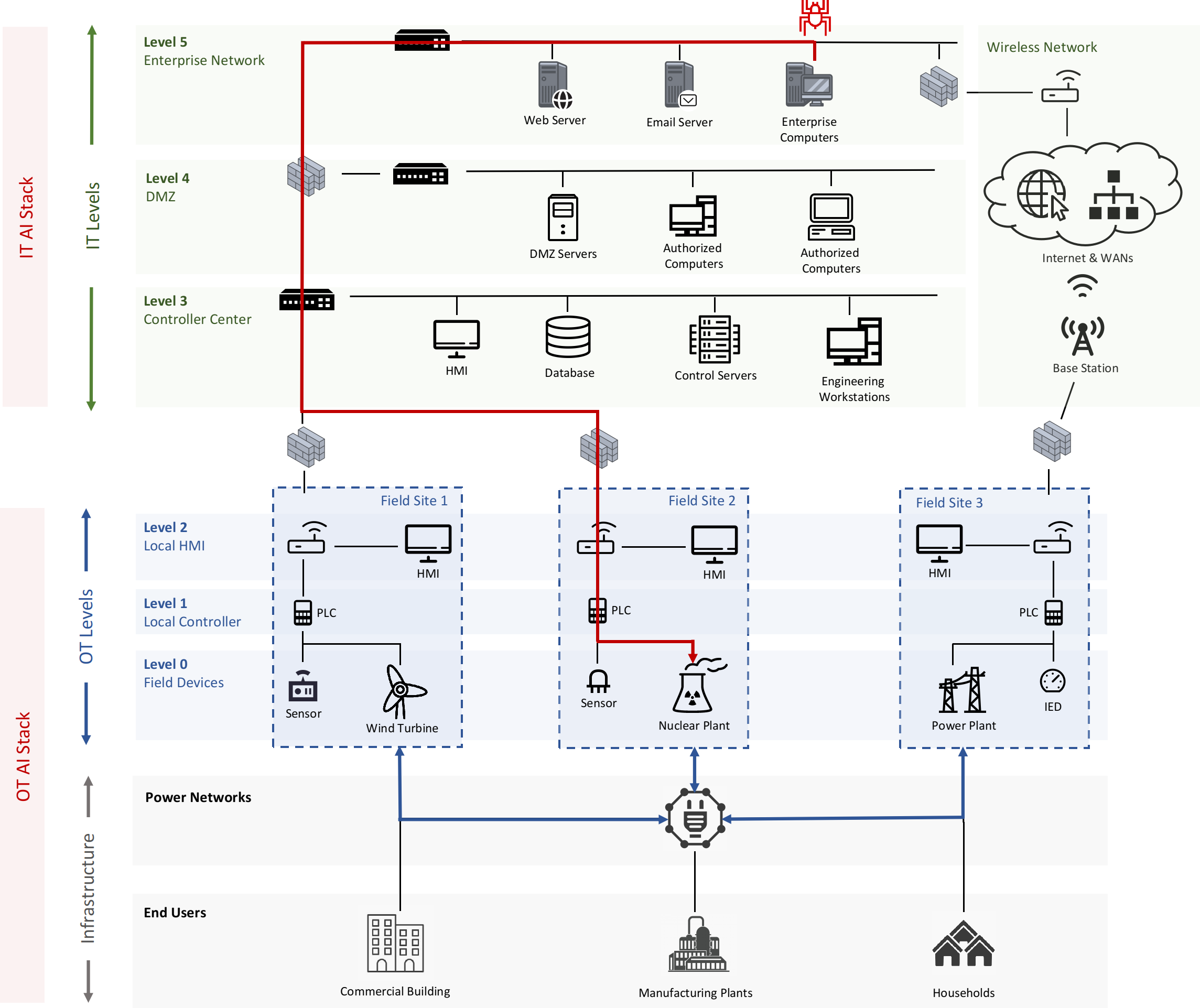}
    \caption{An illustrative example of an ICS network of power systems. Levels 0-2 constitute the OT level for real task execution. Levels 3-5 constitute the IT level for information processing such as scheduling. All components in the ICS cooperate together to deliver electricity to support end users.
    An attacker can follow the red path to attack a physical plant from an enterprise computer. AI stacks in IT and OT levels are introduced to augment the learning capabilities of IT and OT subsystems and improve the overall ICS resilience.}
    \label{fig:ics_illustration}
\end{figure}

Industrial Contol Systems (ICSs) are widely adopted in industrial sectors and critical infrastructures, such as power grids and nuclear plants. They are task-critical implementations of Cyber-Physical Systems (CPSs) to operate and automate the industrial process and consist of numerous subsystems such as SCADA systems and distributed control systems (DCS), field controllers such as Programmable Logic Controllers (PLC) and human-machine interface (HMI), physical plants such as power generators and actuators, and other intellectual electrical devices (IED) \cite{stouffer2011guide,asghar2019cybersecurity,bhamare2020cybersecurity,smidts2019next}. All physical components are organized into a hierarchical, multi-level network and are connected by the fieldbus or wireless networks to perform specific control tasks. Data transmission between levels of the network ensures real-time surveillance for security and control purposes. Fig.~\ref{fig:ics_illustration} shows a notional integrated ICS architecture in the power domain based on the commonly adopted Purdue Model \cite{williams1994purdue}.

An ICS typically consists of several field sites (levels 0-2) that are functional points and can execute real-time tasks. A field site is generally equipped with local PLC controllers and HMIs for automated and human control, some IEDs such as sensors for measurements, and physical plants for specific tasks. Fieldbus connects all components to the cyber components and the controller center.
Examples of field sites include wind turbines and gas power plants.

The fieldbus, such as the CAN bus, is time-critical and fault-tolerant since any time delay and signal distortion can lead to severe faults of field sites. Besides, the fieldbus requires a large communication capacity due to the increasing scale and complexity of contemporary ICSs. For security considerations, the data transported by fieldbus should be filtered by firewalls before sending to other levels.  

The control center (level 3) is a supervisory level designed for real-time system monitoring, control configurations to filed sites, and necessary human operations and interventions. At this level, HMIs provide process information and status to human operators. Control and data servers monitor and adjust control processes and store historical information. Engineering stations provide the developing and debugging environments for control operations. 

The enterprise zone (level 5) is a business network performing general tasks, such as human resource management and internal communications. Data transmission between the controller center and the enterprise zone is restricted and inspected by the demilitarized zone (DMZ) (sometimes called a perimeter network). A DMZ is a subnetwork acting as an intermediary for connected security devices, which is more than a firewall. DMZ ensures that the data can only be transmitted from a high-security level to a low-security level, and the reverse is prohibited. This is because the data in the controller station is more critical and sensitive. The DMZ filters redundant data traffic from the enterprise zone to the controller center that interrupts the control process. The DMZ also adds another layer of security to the ICS. An attacker cannot bypass the DMZ unless he has direct access to the equipment within the DMZ. It prevents malicious access and attacks on the controller and physical plants from cyberspace, providing another layer of security.

The enterprise zone can generally access the Internet or other wide area networks (WANs) (level 5) through another DMZ or a firewall, depending on the security configuration. Other remote field sites can also be connected to the control center via the Internet or WANs from outside. Different security protocols are required for such remote connections.

Following the adaption of the Purdue Model, the network can be further grouped into IT and IT levels based on the functionalities. Levels 0-2 constitute the OT level and are responsible for real-time task execution such as wind turbine control. Levels 3-5 form the IT level and take charge of information processing-related work such as task scheduling and management.
Upon the classic ICS architecture, we introduce AI stacks at IT and OT levels to enable computational intelligence. The AI stack at the IT level enables real-time intelligent cyber detection and response for cyber attacks based on network data, assisting the ICS to achieve cyber resilience. The AI stack at the OT level empowers learning-based methods for controlling physical plants. Compared with traditional model-based control, the AI-enabled learning methods are particularly convenient for OT systems with complex dynamics. The AI stack also facilitates the development of real-time adaptive strategies for anomalies and disturbances. We mention that we do not specifically portray the AI components in Fig.~\ref{fig:ics_illustration} because the AI stack and computational intelligence can be algorithmically built into many   digital devices such as working stations and local controllers. As its scale becomes larger and the structure more complex, the integration of AI stacks benefits the modern-day ICS to operate more adaptively and effectively against cyber attacks and unknown disturbances than traditional ones.

\subsection{Multi-Agent System Framework of Industrial Control Systems}
The classic Purdue model provides a clear visualization of a hierarchical and monolithic ICS architecture. However, it also has some limitations to achieving resilient ICS design as ICSs are becoming more integrated. On the one hand, the growing monolithic ICS architecture makes the network analysis intractable, making it challenging to design effective and resilient control strategies to cope with cyber attacks and external disturbances. On the other hand, more specialized subsystems are integrated into modern-day ICSs so that all components operate in a distributed fashion. 
Therefore, we can decompose the growing ICS into a set of subsystems based on functionalities. The subsystems cooperate to accomplish the overall ICS mission. For example, in the power system, the generating plants work as individual subsystems to produce power, while the delivery grid act as another subsystem to deliver the power to end users. 
Within each subsystem, we also have the cyber-physical architecture. Therefore, the IT and OT division in the Purdue model in Fig.~\ref{fig:ics_illustration} is still applicable for each subsystem. We refer to the divided systems as IT and OT systems, as shown in Fig.~\ref{fig:intro}. The IT and OT systems affect and control the subsystem in a distributed but interdependent fashion. For example, in a generating subsystem, the IT system can monitor real-time information such as generating capacity and output to the grid for power management and coordination. The OT system stabilizes the generating plants and produces power based on the requirements from the IT system.

Therefore, we propose a MAS framework to characterize modern-day ICSs using concepts in distributed control systems. The MAS framework has a two-level interpretation. At the system level, each functional subsystem is regarded as an intelligent agent with a specific objective. All agents interact and cooperate to accomplish the system-wide mission, such as power supply in the infrastructure. At the subsystem level, the IT and OT systems act as intelligent agents, and their interactions lead to achieving the subsystem-specific task objective. For example, a generating subsystem performs a smooth power generation under external requirement changes or malicious attacks on the generating plants. 
The two-level MAS framework is intrinsically interdependent. The interactions at the subsystem level ensure the regular operation of the subsystem and lay the foundation for system-level cooperation. The interaction and coordination at the system level achieve the system-wide ICS mission.

Our MAS framework also enables learning for resilient ICS design and lays the foundation to achieve MAR. On the one hand, every intelligent agent (subsystem and IT or OT system in the subsystem) can utilize network data and learning to adapt to new attacks and disturbances, which provides a flexible and real-time adaption mechanism to achieve resilience. On the other hand, every agent can learn to achieve self-adaption and cooperate with others when the agent has complex dynamics and the interdependencies between agents are not known or planned for. We discuss more details on learning for resilience in Section \ref{sec:learning}.

Our multi-agent perspective aligns with the literature. For example, a hierarchical MAS framework (Hierarchical Multi-agent Dynamic System, HMADS) has been proposed in \cite{rieger2013resilient,rieger2013hierarchical} to characterize resilient control systems. The framework consists of three layers based on high-level functionality: the management layer, the coordination layer, and the execution layer. The management layer provides the high-level system objective and task scheduling. The coordination layer allocates resources to system components to align with the system objective. The execution layer takes charge of controlling and sensing field devices. A subsystem or a controller in the control system is treated as an intelligent agent and is divided into one of three layers based on its sphere of influence. Our framework absorbs the management and coordination layers into the IT, and the execution layer is subsumed under the OT. 

It is clear that a divide-and-conquer approach plays an essential role in the architecture of the large-scale complex ICS. In addition to the focus on IT and OT in this chapter, human technologies (HT) are becoming increasingly prominent. They are central to the resilience of the ICSs as many attacks start with exploiting human error and vulnerabilities. Human-machine technologies have been proposed to augment human performances, e.g., securing human operators from attention failures \cite{huang2022radams}, employees from phishing \cite{huang2022advert}, and users from noncompliance \cite{huang2022zetar,casey2016compliance}. It is natural to extend our framework to include an HT layer to incorporate the emerging human solutions and their interface with IT-OT. 
The hierarchical divide-and-conquer view of ICSs provides a flexible and evolving perspective that enables integrating the systems with new emerging interfaces and technologies. For example, the adoption of IoT technologies and human-machine teaming capabilities can potentially augment ICSs with new cross-layer functionalities and introduce additional dimensions to the ICS architecture.

\subsection{Security Challenges}
The interconnected IT and OT components can create potential vulnerabilities to degrade ICS performance and jeopardize the security of infrastructures. As shown by the red path in Fig.~\ref{fig:ics_illustration}, an attacker can penetrate the ICSs by compromising an enterprise computer and moving laterally within the ICS network to reach the target asset. For instance, an attacker can bring down a nuclear power plant by manipulating the sensors. We have witnessed an increasing number of such attacks, including ransomware attacks, insider threats, and APT attacks, which have raised serious security threats to ICSs \cite{makrakis2021industrial}. We summarize the critical security challenges of ICSs that remain open issues.

First, cascading failures can devastate ICSs due to the multi-level structure and the multi-agent characteristics. The cascading connections of different components provide numerous possible fault-error-failure propagation paths. A single fault of any component, such as hardware flaws or software bugs, can propagate in the network and consequently affect other components, which eventually leads to a system-wide functionality failure. These threats have been noticed and investigated in \cite{zhu2012dynamic,rieger2019industrial}. Cascading failures also make it hard to detect and diagnose the real anomaly in ICSs because of fault propagation. Fixing a single component's error is insufficient to address the general security issues of ICSs. More precise and complex fault detection and identification are required to pinpoint the initial fault and prevent further failures. 

Second, the large-scale connections and heterogeneous components in contemporary ICSs lead to intrinsic complex structures, further increasing the diversity and the complexity of security vulnerabilities. Common security threats have been identified and categorized in \cite{cardenas2011attacks} and precautions can be designed to prevent the existing threats. However, many unknown factors can still cause system-wide malfunctions and jeopardize ICS security. Exploiting and identifying vulnerabilities for modern-day ICSs is still an open project. Besides, the attack surface of subnetworks in ICSs also differs from their characteristics and tasks. For example, generating plants and power grids have different network architectures and security vulnerabilities. The device compatibility is also a security concern in ICSs. It is common to observe obsolete devices in ICSs due to historical reasons. These components are generally not designed for contemporary security considerations and leave potential vulnerabilities. It is challenging to design effective and holistic strategies to make different versions of devices work together and provide enough security guarantees. 

We identify cyber attacks as the third security challenge. A significant feature of contemporary ICSs compared with traditional ones is the adoption of the cyber layer. However, cyber attacks nowadays are causing a broad and profound impact on ICSs. While the cyber layer achieves efficient communications among different components and enables system automation, it also brings more security threats. The existing and recognized cyber threats include communication disruptions and malware attacks such as computer viruses and ransomware. Works such as \cite{uma2013survey,formby2017out} also summarized common cyber attacks and malware in ICSs. 
Although cyber attacks are not originally designed to sabotage ICSs, several recent attacks on ICSs, including the Colonial Pipeline ransomware attack \cite{colonial2021wiki}, have shown their destructive power on infrastructures and economics. Cyber attacks also bring another level of security consideration: intelligent adversaries. Attacks can start malicious and intelligent attacks on ICSs through cyber layers, which differ from the security challenges caused by device faults. Advanced persistent threats (APTs) are an example of intelligent attacks. APT is a cyber attack that aims to penetrate the network and attack a high-value asset. It searches possible attack paths to the target assets stealthily and starts the attack when the time is perfect. APT can avoid existing security defense mechanisms strategically. So traditional methods such as routine security checks may have limited effects on APTs. Although it appears harmless for other components during the network penetration, the attack consequence is devastated because of the loss of the target asset. In short, it is critical to address the cyber threats in ICSs to maintain system resilience.

Human factors also pose security challenges in ICSs. ICSs are integrated with HMIs and workstations for human operators (or experts) to monitor the system status and debug errors. Although the contemporary ICS design keeps improving the human friendliness in operation and facilitating decision support, human operators are still indispensable. For example, human operators are required when processing emergencies such as generating plant breakdown. Besides, human operators are also capable of diagnosing complex errors such as new cyber attacks and faults of specialized devices. Therefore, human experts and their knowledge can benefit ICSs to achieve better performance and resilience. However, security threats also accompany humans and hence jeopardize ICSs.
For example, humans have limited attention. When cascading errors occur, it can be difficult for a human operator to recognize the relevant and irrelevant threats. Also, cyber attacks can launch feint attacks to distract and deceive human attention \cite{hitzel2019art,huang2021radams}. Humans are also prone to be manipulated. There have been phishing scams and espionage activities to steal critical ICS information or plant malware. Some activities can be hard to recognize. Last but not least, insider threats also significantly endanger the security of ICSs. Some of these attacks leverage human operators' access to the core system to steal data or inject malware to sabotage plants. According to empirical research \cite{homeland2016recommended}, social engineering scams and insider threats have become the primary threats to ICSs. Therefore, training and regulating human behaviors to have more efficient and user-friendly ICSs is vital and remains an open challenge.

The security challenges raise severe threats to ICSs such as power grids and nuclear plants, leading to devastating consequences. Therefore, it is critical and indispensable to improve the resilience of ICSs, so that the system can resist malicious attacks and disturbances and maintain an acceptable level of operational normalcy. Traditional threat detection and mitigation are not sufficient to deal with the complexity of modern-day ICSs and the new security challenges. The integration of AI stacks into the ICSs can augment the distributed learning capabilities of the subsystems and improve the overall ICS's resilience.

\section{Resilience of Industrial Control Systems} \label{sec:resilience}
\subsection{Concept of Resilience}
The modern-day ICSs are increasingly integrated with smart devices and services enabled by the Internet of Things and smart automation technologies. This trend is accompanied by a growing attack surface that exposes ICSs to a large number of vulnerabilities across the multiple layers of the system. The traditional security mechanisms, including intrusion detection systems, firewalls, and encryption, are no longer sufficient to protect ICSs from sophisticated and unknown threats. For example, Advanced Persistent Threats (APTs) are one class of threats that can stay in the system for a long time and stealthily reach the target resource. In addition, many emerging attacks (e.g., recent Log4j attacks \cite{log4j2021mitre} and supply chain attacks \cite{kshetri2022economics}) and unknown ones have made it harder or impossible to prepare the system ahead of time and achieve perfect security. Instead, resilience is a critical aspect of cyber protection that provides a supplemental means to mitigate their impact once the traditional methods fail to thwart them. Resilience plays an even more important role in defending against unknown or unanticipated attacks as it becomes the last resort to safeguard ICSs from calamitous collapse. 

\begin{figure}
    \centering
    \includegraphics[scale=0.14]{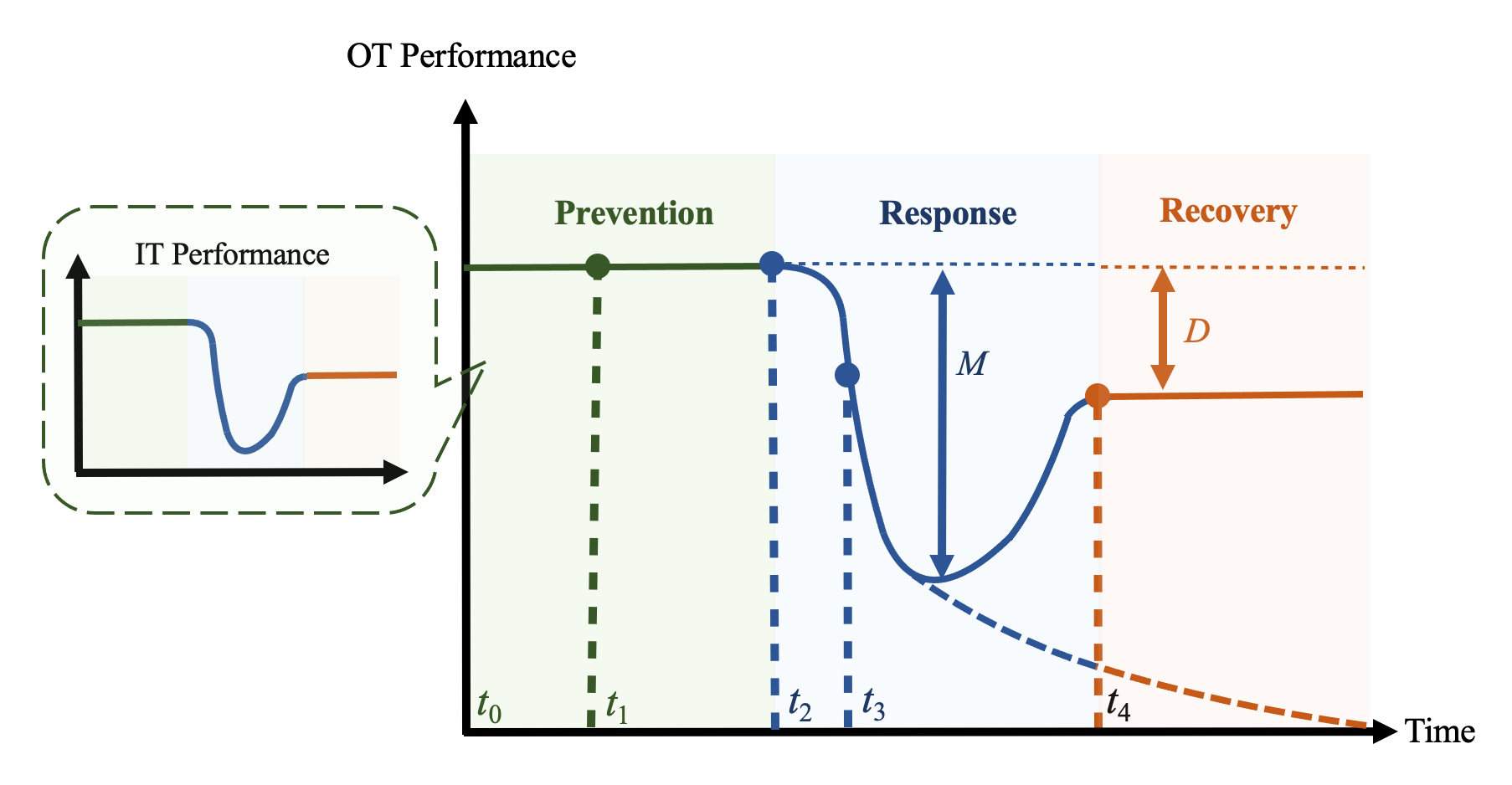}
    \caption{The resilience of an ICS can be measured by the OT-level performance. A resilient ICS experiences three stages after an attack is launched at $t_1$, including prevention, response, and recovery. A $(T,D)$-resilient ICS to the attack can agily respond to the attack by recovering within $T$ units of time and maintaining a loss of performance $D$. The IT-level performance is subsumed into the OT-level one. The IT-level performance determines the success rate and the timing of an attack.}
    \label{fig:resilience_performance}
\end{figure}

An ICS is a large-scale system of systems. Its resilience is a composition of the resilience of many interacting subsystems. Each subsystem consists of its IT and OT units. Hence the resilience of the system relies on the resilience of IT and OT. For a given system, subsystem, or unit, its resilience is defined by its ability to respond to an attack, maintain its function after the attack, and recover from the attack. Illustrated in Fig.~\ref{fig:resilience_performance}, we divide the operation into three phases. The first phase is the prevention, or known as the ante impetum stage, where less sophisticated attacks are thwarted while some attacks launched at time $t_1$ succeed in bypassing the defense and starting to navigate within the system at $t_2$. The second phase is the interim impetum stage (response). The attack breaches the security defense and aims to look for and take over the control of the targeted resources. 
For example, in the infamous Target data breach \cite{plachkinova2018security}, the APT attacker, after infiltrating the network by leveraging the third-party vendor’s security, moves laterally from less sensitive areas of Target’s network to areas storing consumer data. A resilient mechanism at this stage aims to respond to this breach and deter the attacker from moving forward to reach the target. Without it, the system performance will progressively deteriorate, and the system will eventually break down. In contrast, an appropriate resilient mechanism detects the breach at $t_3$ and immediately foils the attack by taking appropriate measures (e.g., disconnecting the subsystem, rebooting the system, or triggering a new set of authentication rules) to mitigate the risk. The third stage is the post impetum stage (recovery), where the defender can remove the attacker from the system and recover the compromised system to its ante impetum state or maintain an acceptable performance at the loss of the performance level $D$. Similar concepts on resilience are also discussed in \cite{mcjunkin2017electricity}.

\subsubsection{Metrics for Resilience}
For a given attack, we say that an ICS is $(T,D)$-resilient if the system can recover from the attack within $T$ units of time and maintain a maximum loss of performance $D$. Here, $T$ is a measure of how quickly the response is. From Fig.~\ref{fig:resilience_performance}, $T=t_4-t_2$ can be further viewed as the duration of the response stage. The performance loss $D$ quantifies the post impetum performance. A non-resilient system goes through an irreversible performance degradation during the operation, as depicted counterfactual-wise in Fig.~\ref{fig:resilience_performance}.

The concept of $(T,D)$-resilience provides basic metrics for ICSs. Another commonly used metric for resilience is the measure of the total performance loss after an attack. The maximum magnitude of performance loss $M$ at the interim impetum stage can also serve as an indicator of resilience. Note that the metrics discussed above build on the performance measures of the OT system or the operation at the OT level. It is important to recognize that the defense effort at the prevention stage in the IT system has a significant impact on the metrics even though it is not directly measured, and it is quantified through the OT performance. The effort to thwart the attack, such as the deployment of honeypots \cite{huang2019honeypot}, moving target defenses \cite{zhu2013game}, and zero-trust mechanisms \cite{rose2020zero}, creates more difficulty for the attacker to be successful, thus reducing the rate of successful attacks and the impact of their consequences. In Fig.~\ref{fig:resilience_performance}, this effort can be roughly measured by the time that it takes for an attack in the system to compromise the targeted OT assets. Many recent works on cyber resilience focus on resilience at this stage \cite{segovia2020cyber,mertoguno2019physics}. We can zoom into the IT system and further quantify the cyber resilience of an ICS. For example, in \cite{kerman2020implementing}, a zero-trust cyber resilient mechanism is developed for an enterprise network. The trustworthiness of each user is measured over time to disrupt unknown insiders from lateral movement. The resilience of the IT system is measured by the time it takes to deter an attack from the target asset and the distance between the attacker and the asset. Fig.~\ref{fig:resilience_performance} provides a resilience measure of ICSs through the OT performance which consolidates the impact and the performance of the IT system. We can also zoom into the prevention stage and create a similar performance measure for the IT system. 

\subsubsection{Comparison with Related Concepts}
\begin{figure}
    \centering
    \includegraphics[scale=0.14]{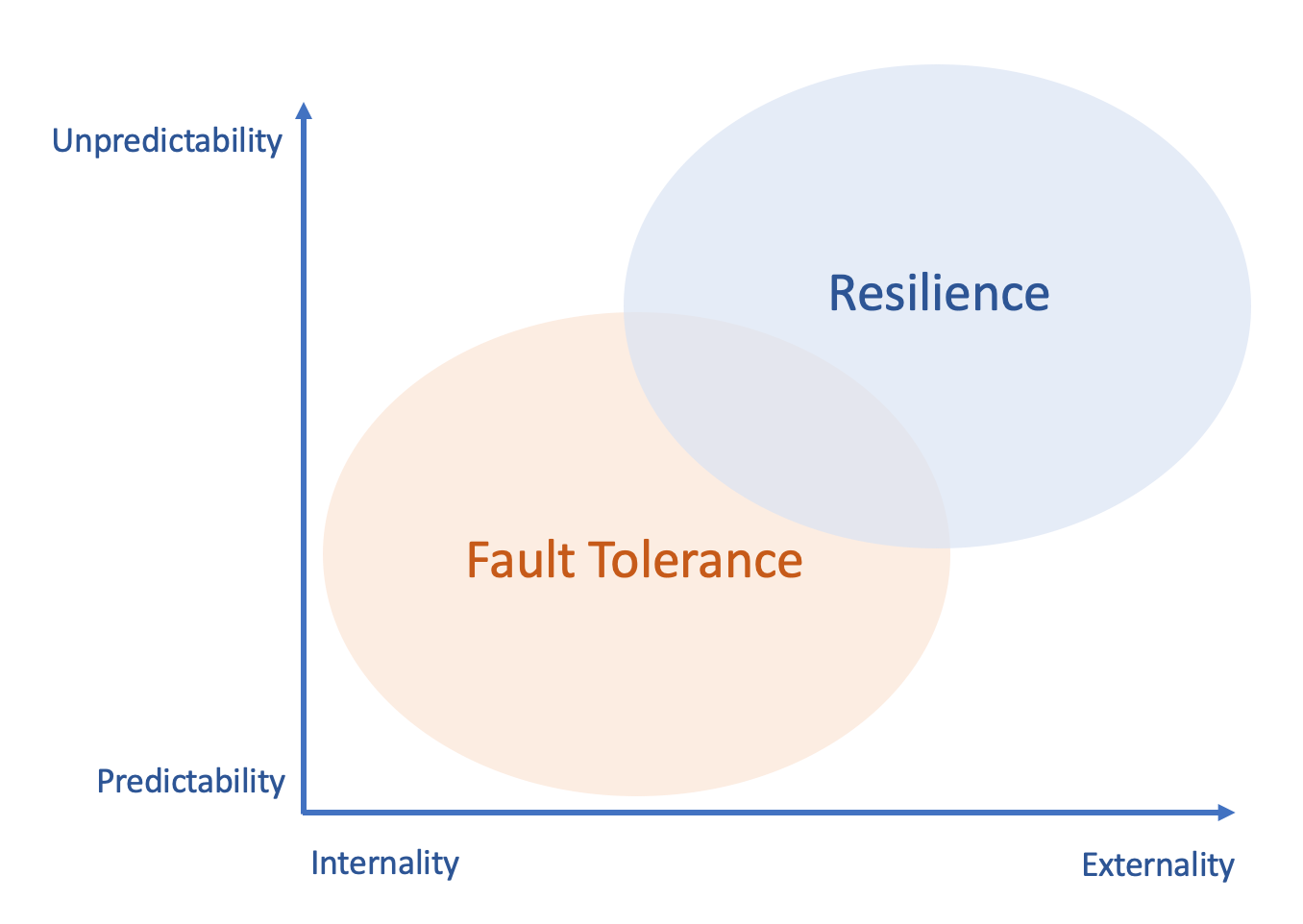}
    \caption{The distinction between fault tolerance and resilience from the dimensions of unpredictability and the externality of the events. The two concepts overlap for some events.}
    \label{fig:resilience_fault_tolerance}
\end{figure}

There are subtle differences between resilience and related concepts, such as fault-tolerance, robustness, and security. Fault tolerance is also a system property that aims to enable the system to maintain its core functions in the event of failures. Illustrated in Fig.~\ref{fig:resilience_fault_tolerance}, fault tolerance often prepares for anticipated failures, and those happen internally caused by fatigue, corrosion, or manufacturing flaws. Resilience, in contrast, deals with unpredictable ones and those caused by external influences, such as cyber-attacks, natural disasters, and terrorism. The event that the system aims to prepare for distinguishes the type of system properties that we need to focus on. 

Robustness is a well-known system concept for control systems. The goal of robustness is to design systems to withstand a set of prescribed uncertainties or events. Robustness prepares for them and enables the system to function when they occur, which is beneficial for resilience planning. However, it does not focus on preparing the system for recovering from an unknown event that has disrupted the system. It is often either cost-prohibitive or nonviable to design systems robust to all possible events while maintaining their functions. 
Robustness and resilience are complementary to each other. The low-probability events that are costly for robustness should be prepared by resilient mechanisms while the high-probability events, such as thermal noise and load variations, should be handled through robust designs. The concept of robustness is often used for OT, e.g., controller designs and operation planning. 

Security is a related concept that focuses on the prevention of adversarial events to safeguard the confidentiality, integrity, and availability of IT systems. Despite the effort to create a cyber defense to detect and foil attacks from penetrating IT, there is no perfect security that can assure the system of no attacks. In contrast, cyber resilience plays a significant role that enables the IT and OT systems to function even though the attack is in the network already.

\subsection{Resilience of Distributed Systems}

\begin{figure}[h]
    \centering
    \includegraphics[scale=0.23]{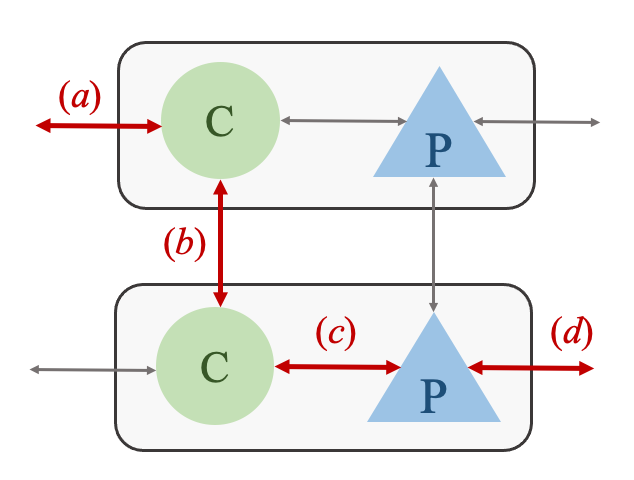}
    \caption{Some illustrative attacks on subsystems of an ICS. The connectivities among subsystems increase the attack surface. An attacker can move laterally through the IT network from (a) to (b) and locates the OT asset using (c). The attack can cause cascading failures through (d).}
    \label{fig:attack_ics}
\end{figure}

ICS are large-scale systems. Modern-day power grids are composed of many distributed energy resources at the edge of the power grid. The increasing interdependencies and connectivities among subsystems have exposed the ICS to a large attack surface. As illustrated in Fig.~\ref{fig:attack_ics}, an attacker can first gain the privilege of an IT subsystem through (a) and then move laterally within the IT network to find the target asset through (b). The attacker takes down the OT asset of the targeted subsystem, and its compromise leads to cascading failures in the OT system. This type of incident has been witnessed in the Ukrainian power grid attack \cite{case2016analysis} and the oil pipeline system attack \cite{colonial2021wiki}.
They both are attributed to APTs, which carry out a kill chain consisting of a sequence of resourceful, stealthy, and strategic attacks to reach their target.  

\begin{figure}
    \centering
    \includegraphics[scale=0.31]{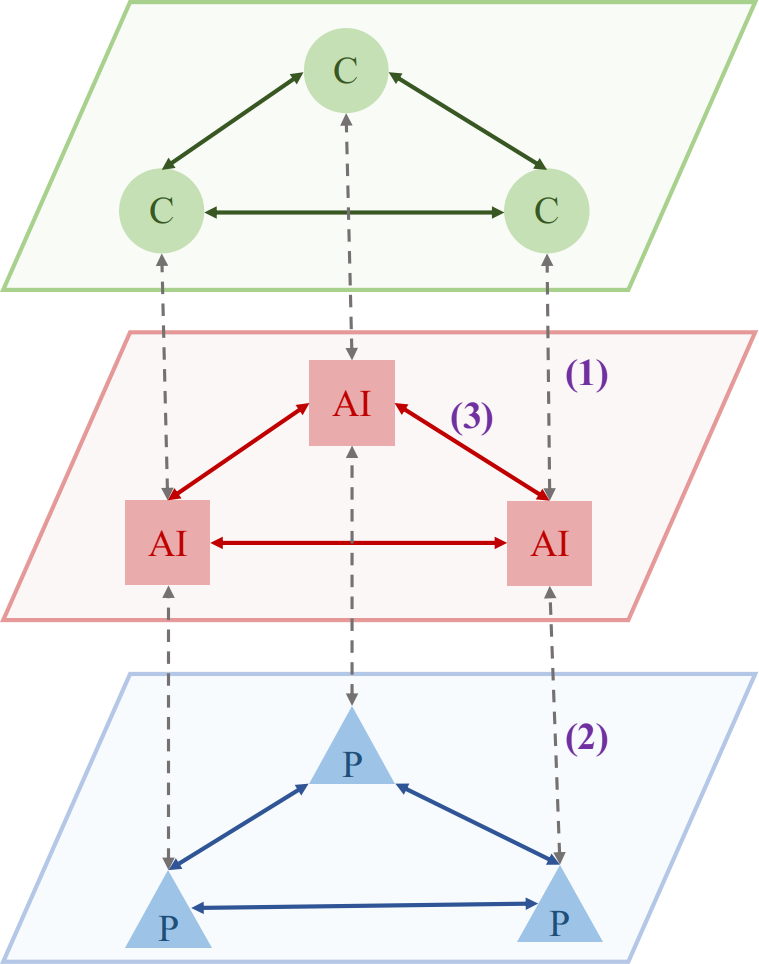}
    \caption{An illustration of interactions between AI, IT, and OT agents of three subsystems. The interactions include (1) interface between a C agent and an AI agent, (2) interface between a P agent and an AI agent, (3) interface between two AI agents in two subsystems. 
    The C and P agents from different subsystems can also be interdependent. The AI agents interact with C and P agents in the same subsystem and other subsystems to achieve distributed and collaborative resilience.
    }
    \label{fig:ai_stack}
\end{figure}

As the attackers are growingly sophisticated, resilience becomes increasingly essential for ICSs. However, it is challenging for large-scale systems through centralized control and operations. One-point failure can propagate to the entire system, and a centralized operation is less flexible as the entire system needs to be reconfigured for the failure. 
Distributed control, on the other hand, enables the agility of the entire system. The system can still maintain its core functions well even when a subset of subsystems fails. In addition, as the decisions are distributed, it will be more convenient and faster for each subsystem to respond to the disruption and its connected subsystems. The challenge, however, to achieve it is the need for machine intelligence to enable detection and response through fast data analytics and decision-making. 
Hence, to enable distributed and collaborative resilience, we need to create an AI stack between the IT and OT systems of each subsystem. Illustrated in Fig.~\ref{fig:ai_stack}, AI agents are introduced to connect cyber IT agents (C) and physical OT agents (P)\footnote{Here we call the IT and OT systems as agents to align with the names in our MAS framework. AI agents refer to the introduced AI components such as computational devices.}. 

\begin{figure}
    \centering
    \includegraphics[scale=0.22]{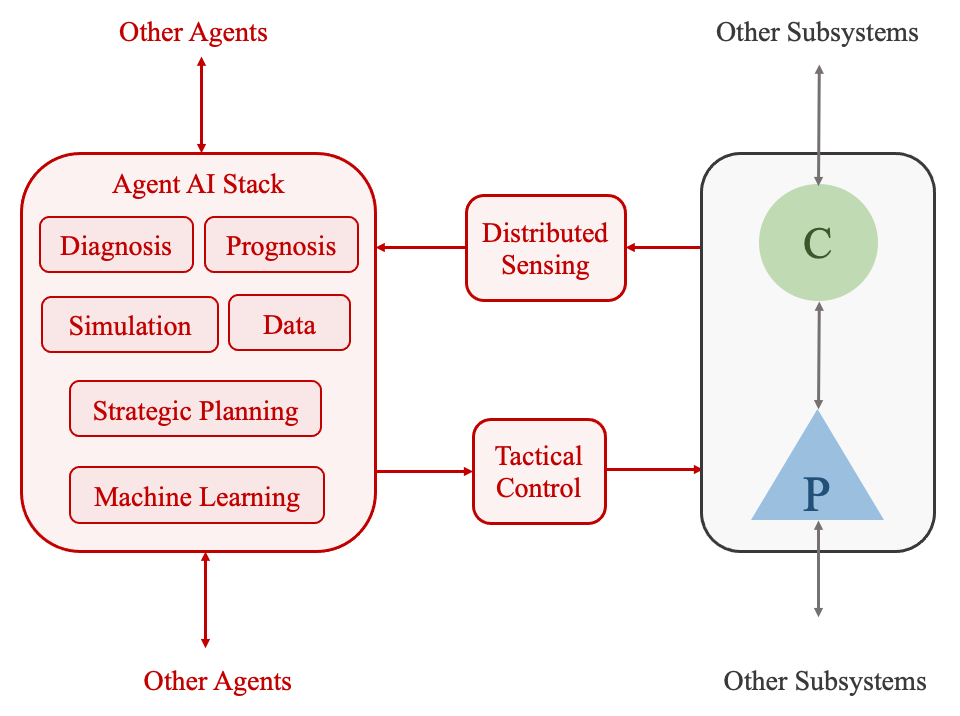}
    \caption{An illustrative functional diagram of an AI-augmented subsystem. An AI agent of a subsystem monitors the behaviors of its C and P agents and communicates with other AI agents to create tactical control and planning decisions to respond to malicious attacks and disturbances.}
    \label{fig:ai_agent}
\end{figure}

A more detailed and functional diagram of an AI-augmented subsystem is also demonstrated in Fig.~\ref{fig:ai_agent}. An AI agent serves as an interface between a C agent and a P agent and is equipped with necessary functional blocks, including diagnosis, prognosis, simulation, data, strategic planning, and machine learning, to create fast online responses to an unanticipated event. The AI agent can also interact with other connected AI agents. This architecture enables the AI agent to control the performance of its associated subsystem as well as coordinate with other subsystems. 
When a cyber-attack occurs locally on the subsystem, the AI agent can monitor the behaviors of the C and P agents in its subsystem and make tactical control and planning decisions to respond to the event. Each subsystem can achieve its own resilience in a distributed fashion. We refer to this type of resilient mechanism for the large-scale ICS as distributed resilience. As the failures can propagate across multiple subsystems, the AI agents need to communicate and coordinate to achieve collaborative resilience. Sharing of information and intelligence among AI agents can not only improve the distributed resilience of their subsystem but also enhance the global resilience of the large-scale system. For example, in \cite{achbarou2018new}, we have seen that the sharing of information in intrusion detection systems can help the entire network defend against zero-day attacks. In \cite{sun2021data}, the collaborations among multiple machine learners have led to significant improvement in the learning as well as resilience to data poisoning attacks. 

\section{Learning and AI for Resilient ICSs} \label{sec:learning}
\begin{figure}
    \centering
    \includegraphics[scale=0.17]{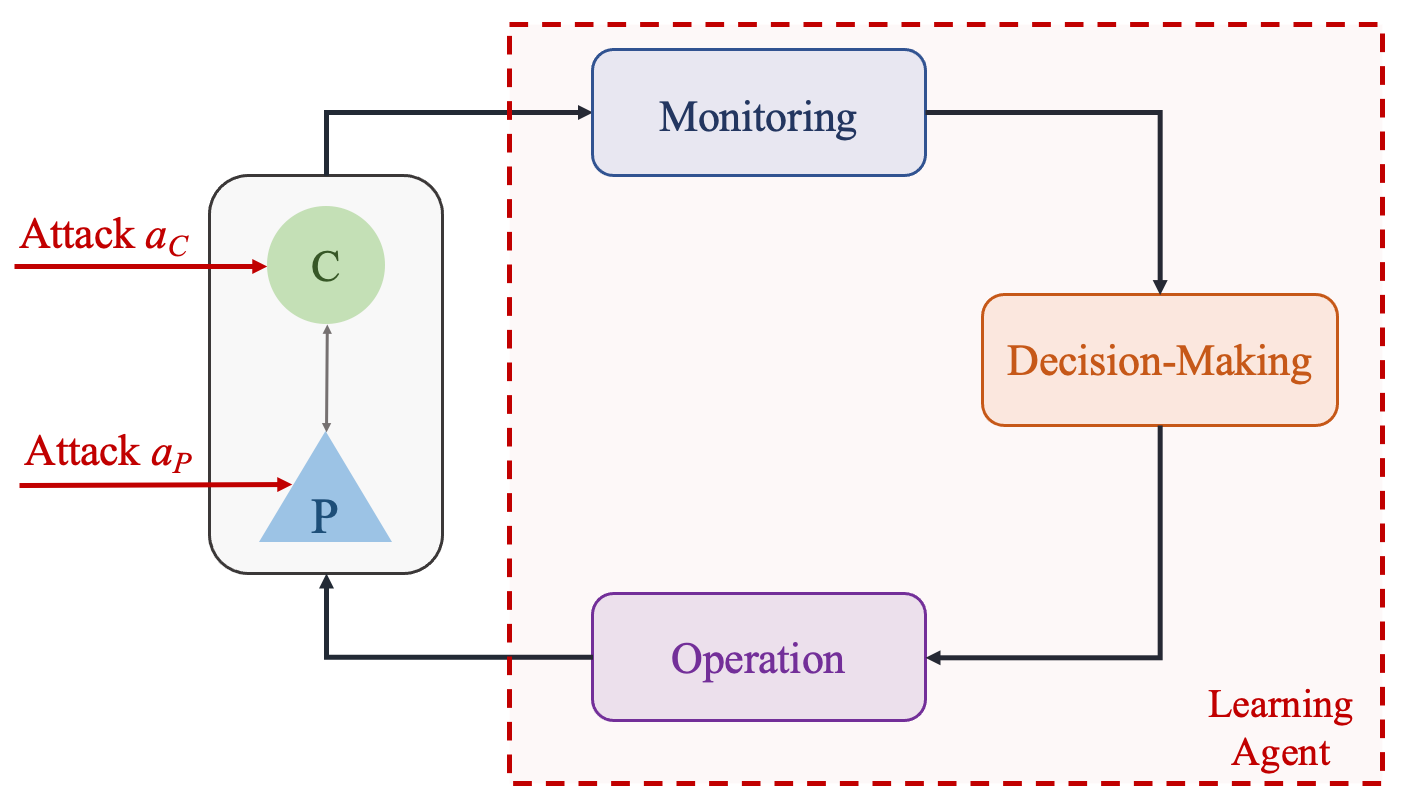}
    \caption{The feedback structure of an AI learning agent in an ICS subsystem, which consists of three major components: monitoring, decision-making, and operation. The learning takes different forms for attacks on the IT and OT systems of the subsystem.}
    \label{fig:ics_learning}
\end{figure}

Learning plays an essential role in achieving resilience since the attacker's identity and behaviors can be unknown to the system. Learning creates a constant adaptation to new changes and uncertainties in the ICS as well as its subsystems. Feedback is an essential architecture of the learning at the AI stack as we have seen in \cite{huang2022reinforcement}. Fig.~\ref{fig:ics_learning} illustrates a baseline structure for learning in each subsystem of the ICS, which consists of three key modules, i.e., monitoring, decision-making, and operation. Each module takes a distinct form for attacks on IT and OT systems. We discuss how the learning helps build resilient ICSs at the subsystem and system levels and review the related learning methodologies for resilience.

\subsection{Learning in Cyber IT System}
In the IT system, a learning agent monitors the cyber behaviors of the users and the network to detect the anomaly and adversarial behaviors. The goal of the IT operation is to mitigate the impact of the attack, which can propagate to the OT if not carefully handled. The learning process in this system depends on the sophistication of attacks and resources of the decision-maker, including his knowledge, computation, and maneuverability. 
Based on the functionality, we can summarize the components in the feedback loop in Fig.~\ref{fig:ics_learning} into two modules: cyber detection and cyber mitigation. Two modules work jointly to enable cyber resilience.

\subsubsection{Learning for Cyber Detection}
Cyber detection is fundamental for cyber resilience. An AI agent can learn from cyber data such as network traffic and log traces to detect and recognize cyber threats, including internal anomalies and external attacks on cyber agents and human operators. Based on the learning pattern, the learning methods for cyber detection can be categorized as supervised learning, unsupervised learning, and semi-supervised learning \cite{mitchell2014survey}. Each type of learning provides different advantages to achieving successful detection.

\emph{Supervised learning} uses cyber data and labeled system behaviors (for example, normal or abnormal) to train a cyber detection model offline and perform detection online. 
Many works have developed successful learning-based detection models based on different supervised learning techniques. For example, 
in \cite{linda2009neural}, Linda et al. used neural networks (NNs) to detect cyber intrusions for critical infrastructures. 
Their neural network model was trained by actual network data and showed a perfect detection rate and zero false positives on testing cyber attacks. 
In \cite{wang2019detection}, Wang et al. adopted random forest (RF) to detect natural disturbances and artificial cyber attacks for power systems. The model was trained by historical network data and relevant log information. The experiments on an open-source simulated power system dataset reported a successful detection rate higher than 93\%.
Mokhtari et al. in \cite{mokhtari2021machine} used $K$-nearest neighbors (KNNs) and decision tree classifier (DTC) as a tool for cyber detection. The data collected by the SCADA system were used for detection model training, and their reported detection precision and accuracy were higher than 97\%. 
Many works such as \cite{beaver2013evaluation,hink2014machine,ozay2015machine} also made a thorough comparison of different supervised learning methods for cyber detection and demonstrated the effectiveness of the learning-based methods. 
Supervised learning is also used for human-related attack detection. In the work of \cite{alam2020phishing}, an RF and DTC-based model was developed to detect phishing attacks and help human operators stay safe. An NN-based detection model was also proposed in \cite{abdelaty2021daics} to reduce false alarms in ICSs to ease human operators' cognitive load.

Supervised learning enjoys a high cyber detection precision and accuracy as we have observed in the literature. However, an AI agent can hardly detect new anomalies and attacks with supervised learning because of missing labels in the dataset. Unless updating detection models constantly, an AI agent may not detect zero-day attacks and fail to provide cyber resilience for the system. Besides, acquiring labeled cyber data is often a challenge.
As a complementary approach, \emph{unsupervised learning} can handle unlabeled cyber data and provide more generic detection. The main idea behind unsupervised learning is feature-based clustering. A learning agent first identifies the features of cyber data and then searches for potential anomalies or attacks by clustering the featured data. Many recent works have focused on developing detection tools with unsupervised learning.
For example, Maglaras and Jiang in \cite{maglaras2014intrusion} adopted the One-Class Support Vector Machine (OCSVM) to detect malicious network traffic in SCADA systems. The detector extracted attributes from network traces and then was trained to classify the attacked nodes that jammed the network traffic.
Alves et al. in \cite{alves2018embedding} used K-means clustering to detect data interception, data injection attacks, and denial of service (DoS) in industrial PLC networks. The anomaly classification and learning were based on the features of real-time data streams, such as pack latency and processing information. 
Kiss et al. in \cite{kiss2014data} investigated detection in critical infrastructures based on K-means and MapReduce paradigm to achieve real-time cyber attack detection.
To address the anomalous behavior detection of building energy management systems, Wijayasekara et al. in \cite{linda2012computational,wijayasekara2014mining} developed an efficient detection method based on the modified nearest neighbor clustering and fuzzy logic, which not only performed faster than traditional alarm-based detection methods, but also was able to provide linguistic interpretation of the identified anomalies.

Unsupervised learning may produce more false alarms in detection due to a lack of labels compared with supervised learning \cite{umer2022machine}. Human intervention for diagnosis is useful, but it may delay the IT operation and result in less cyber resilience in the IT system. To overcome the issue, many works consolidate supervised and unsupervised learning to create \emph{semi-supervised learning} for effective cyber detection. 
In recent work \cite{kravchik2018detecting}, Kravchik and Shabtai developed a window-based cyber attack detection using convolutional neural networks (CNN) for critical water infrastructures. The CNN was trained to predict normal system behaviors, and anomalies were detected every time window based on the error between predicted and observed system behavior. 
Marino et al. in \cite{marino2019cyber} adopted unsupervised learning for ICS anomaly detection classification and evaluated its performance with supervised learning. The combination of the two learning methods empowered the detection of unseen cyber anomalies and improved the detection accuracy.

We refer the readers to recent surveys \cite{umer2022machine,bhamare2020cybersecurity,anthi2021adversarial,handa2019machine} for details on machine learning algorithms for cyber detection in ICSs.

\subsubsection{Learning for Cyber Mitigation}
The cyber mitigation module serves as the decision-making and operation units in the cyber learning architecture and is the key to guaranteeing cyber resilience. However, learning for cyber mitigation is more challenging because the mitigation strategies rely on tasks and network structures. Traditional cyber mitigation strategies do not pay attention to the higher-order impact of the mitigation strategies on the system performance and the complex roles of attackers who can strategically and stealthily counteract the system strategies. To address these challenges, game-theoretic learning and reinforcement learning (RL) are used for effective and proactive cyber mitigation.

\emph{Game-theoretic learning} treats the attacker as a rational player and learns a strategic solution by observing the attacker's attack trace. It provides a suitable framework to develop attack-aware solutions to combat APTs in the IT system, where attackers aim to penetrate the IT system to attack the target asset \cite{rubio2019current}. For example, Huang and Zhu in \cite{huang2018analysis,huang2019adaptive,huang2020dynamic} proposed a Bayesian learning mechanism to defend APTs in the cyber system proactively. They used a multi-stage dynamic Bayesian game to characterize the long-term interaction between a stealthy attacker and a proactive defender. The Bayesian learning was used to update the defender's knowledge of the attacker along the interaction and assisted the defender in developing effective defense strategies for cyber resilience.
Game-theoretic learning has also been used in defensive deception to achieve effective cyber mitigation. A learning agent can proactively learn to deceive the attacker with false targets or steer the attacker to a non-hazard zone to mitigate attack consequences \cite{zhang2020game,huang2020strategic,zhu2021survey,pawlick2021game}.

A \emph{reinforcement learning} agent seeks a strategy that minimizes the attack consequence over some time from observed data. For example, in \cite{huang2019honeypot}, the network defender used RL to adaptively allocate honeypot resources to trap the attacker in the target honeypot for as long as possible.
RL has also been shown to be effective for human-related cyber attacks such as phishing and attention manipulation. For example, Huang and Zhu in \cite{huang2021advert} proposed a phishing prevention mechanism (ADVERT) using RL to generate adaptive visual aids to counteract inattention and improve the human recognition of phishing attacks. Experiments with human volunteers showed an accuracy improvement of phishing recognition from 74\% to a minimum of 86\%. Another human attention management mechanism (RADAMS) using RL was proposed in \cite{huang2021radams} to help human operators fight against informational attacks, such as fake alerts. The proposed mechanism used RL to de-emphasize alerts selectively and reduce human operators' cognitive load. Experimental results showed that the developed attention management mechanism reduced the risk by as much as 20\% compared with default strategies.

\subsection{Learning in Physical OT System}
In the OT system, the goal of a learning agent is to create a learning-based resilient control mechanism that can further mitigate the attack impact on the OT system when the attacker reaches the OT \cite{homeland2016recommended}. The OT learning mechanism also follows the feedback architecture depicted in Fig.~\ref{fig:ics_learning}. The monitoring of the OT assets, including sensors, actuators, and terminal units, creates situational awareness for the decision-makers to adaptively reconfigure the control system to threats and environmental changes.
The learning agent can either learn the uncertainties first and then make decisions or directly learn the control decisions. 

Unlike the OT system, physical plants in the OT system resort to physical models to operate normally. The controller design for physical agents is also model-specific. Therefore, effective and resilient control in the OT system relies on the physical models.
However, exact physical models can be hard or impossible to obtain as contemporary ICSs become more integrated, making resilient control even more challenging. Therefore, many research resorts supervised learning to estimate the physical dynamics and then develop adaptive control strategies for possible OT system attacks. For example, Li and Zhao in \cite{li2021resilient} studied learning-based resilient and adaptive control strategies for uncertain deception attacks in CPSs. The neural network was used to approximate unknown nonlinear and switching physical system dynamics. Then a dynamic surface-based adaptive controller was designed for sensor and actuator deception attacks. The resilient controller was validated on continuously stirred tank reactor systems.
Liu et al. in \cite{liu2021adaptive} proposed observer-based adaptive neural network control for nonlinear CPSs subject to false data injection attacks. The neural network was used to adaptively approximate the unknown nonlinear functions of the CPS. Then a resilient feedback controller was designed to ensure bounded output under malicious attacks. 
%
The model-free adaptive controller design for unknown dynamics was also studied in \cite{wang2017neural,farivar2019artificial}. Both work used supervised learning and neural network models to develop adaptive control strategies to cope with sensor faults and malicious attacks. The system resilience was guaranteed by the stabilizable controller and bounded system output.


Supervised learning has been useful for predicting physical models. However, dealing with malicious and intelligent attacks in the OT system is insufficient. To address the issue, game-theoretic learning is used to find control strategies for physical resilience because of its ability to characterize strategic interactions with adversaries. A learning agent learns to reconfigure the control system by anticipating the attacker's behavior. An example of game-theoretic learning in the OT system is moving target defense (MTD) in ICSs. The defender protects a multi-layer network system and prevents the attacker's penetration by changing the network configuration of each layer. The attacker exploits each layer's vulnerabilities and tries to penetrate the entire system. Zhu and Ba{\c{s}}ar in \cite{zhu2013game} investigated this issue by assuming two players have no knowledge of each other. They proposed an iterative game-theoretic learning algorithm for both attacker and defender to collect more information about each other and a develop better attack and defense strategy. The equilibrium solution was used as the protection plan. The work \cite{sengupta2020multi} also studied MTD strategy using game-theoretic learning in a Bayesian Stackelberg game setting.

\subsection{Cyber-Physical Co-Learning in Subsystems}
\begin{figure}[h]
    \centering
    \includegraphics[scale=0.5]{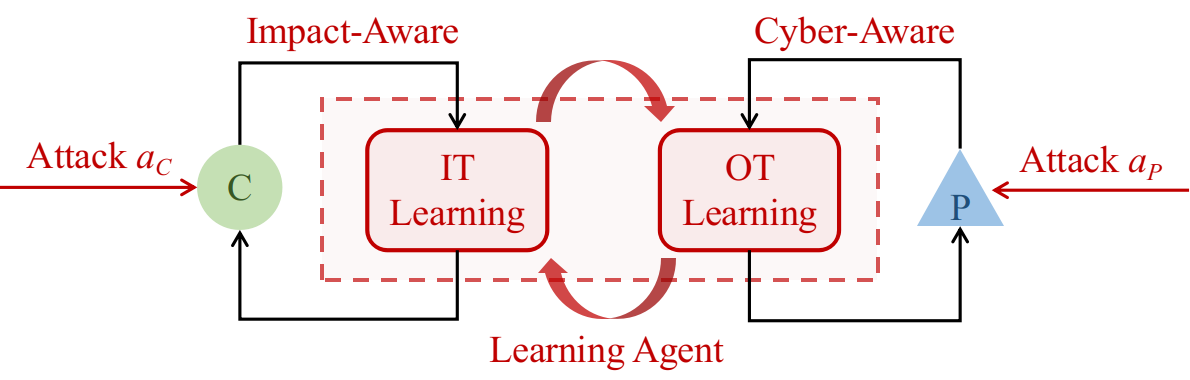}
    \caption{Cyber-physical learning is a co-learning process between the IT and the OT systems. A learning agent consists of IT and OT learning modules. The attacks on the C and the P agents are denoted as $a_C$ and $a_P$. 
    IT learning needs to be aware of the future impact of the learning mechanism; it creates an adaptive mechanism to prevent the propagation of attack $a_C$ to the OT system with the best effort. OT learning needs to consider that IT learning can be imperfect; it creates a cyber-aware mechanism resilient to $a_P$.}
    \label{fig:cross_layer}
\end{figure}

Commonly, the learning modules for IT and OT in a subsystem are designed separately. Each module monitors or senses measurements in its system and operates or controls units in the associated layer. Since the IT and OT performances are interdependent, as illustrated in Fig.~\ref{fig:cross_layer}, it is necessary to create co-learning mechanisms that allow the two learning modules to interact with each other. The principles of cross-system CPS coordination designs in the monograph \cite{zhu2020cross} provide initial attempts to develop effective co-learning mechanisms. 

We discuss how the principles can be applied to co-learning designs. On the IT side, apart from learning to reconfigure IT systems to cope with cyber attacks, the IT learning also needs to consider the impact of cyber learning mechanism to the OT performance. For example, in the learning-based honeypots that aim to engage an attacker to learn his behaviors, the reinforcement learning methods used in \cite{huang2019honeypot} adaptively reconfigure the honeypots by trading off between the reward and the OT related cost. The reward is quantified by the information garnered from the the attacker's behaviors, while the cost arises from the possibility that the attacker can abscond from the honeypot and enter the production IT system which is connected with the OT, and it is relate to the OT performance. It is essential for the learning agent to consider the OT-level consequences while learning for that honeypot defense strategies. In this way, the IT system makes more effective decisions and aligns its perception of the consequences with the goal of the entire system. 

Similarly, on the OT side, the OT learning aims to learn the effective controls to stabilize the OT system and maintain the OT performance while considering possible consequences from the IT system (e.g., the probability of the attack penetrating the IT and the capability of the attacker). The knowledge sharing enabled by the learning agent provides the OT learning with appropriate countermeasures, expedites the learning process, and reduces uncertainties about the anticipated outcomes. 
For example, the OT learning can learn to switch between controllers to cope with different situations and achieve OT resilience \cite{xu2015secure}. One offline but optimal controller is designed to guarantee a robust performance to anticipated uncertainties or attacks. The other online but simple controller can stabilize the system. When the OT system is attacked (e.g., sensors or actuators are compromised), switching from an optimal controller that relies on compromised sensors or actuators to a simple controller that can maintain the basic functions of the control system is key to OT resilience. Learning when to switch from one controller to the other relies on attack detection and the prior knowledge or inputs from the IT system. 

We provide two baseline frameworks of cyber-physical co-designs, based on which we can further build co-learning algorithms. The work \cite{xu2018cross} proposed a resilient cyber-physical design to mitigate cyber attacks to industrial robotic control systems. A partially observed Markov decision process (POMDP) was used to model the IT behavior of the control system, and a time-delay dynamical system was adopted to characterize the OT behavior. A cyber state that could affect the OT dynamics was specified to represent the impact from the IT to the OT. A threshold-based control strategy was developed to achieve a secure and resilient mechanism by jointly considering the interdependence of the two systems. 
The other example is the secure and resilient cyber-physical framework of multi-agent systems proposed in \cite{xu2015cyberphysical}. A signaling game was used to capture the communications between IT devices in adversarial environments. The OT control design was formulated as an optimal control problem. The holistic framework was established by an integrative game between IT and OT systems achieve secure and resilient control.

The cyber-physical co-learning ensures each subsystem in the ICS operates normally against cyber attacks and disturbances. The learning-based cyber detection and mitigation enable a flexible reconfiguration in the cyber system to retard attack penetration and protect OT assets. The learning-based control facilities the design of resilient control for complex OT systems. The information and knowledge sharing in the cyber-physical co-learning allows the IT and OT systems to be aware of and adapt to each other during the learning so that holistic and effective IT-OT strategies can be learned to achieve subsystem-level resilience.

\subsection{Multi-Agent Learning}
The cyber-physical co-learning enables subsystem-level resilience and is the foundation to achieving the system-level resilience and MAR. At the system level, functional subsystems in the ICS act as intelligent agents and cooperate to accomplish system-wide missions. Therefore, the overall resilience of the ICS is built upon the resilience of its subsystem. However, the ICS resilience is not a mere replica of one resilient subsystem due to the subsystems' diverse functionality and complex interdependencies. The resilience of one subsystem also affects the resilience of others. The type of coordination between the IT and OT learning of one subsystem is also needed between two subsystems. To this end, the learning structure depicted in Fig.~\ref{fig:ics_learning} can be extended to a multi-agent learning framework.

Compared with centralized architectures, the multi-agent communications and interactions in ICSs enable multiple subsystems to exchange information and improve their situational awareness and prepare for forthcoming failures. For example in \cite{chen2019control}, a feedback-based adaptive, self-configurable, and resilient framework for the unmanned aerial vehicle (UAV) overlay network has been proposed to provide coverage to underlay devices to maintain interconnectivity despite adversarial behaviors that can disrupt the network. Each agent senses the information of its connecting agents and reaches a coordinated mechanism to achieve the coverage goal. It has also been observed in \cite{chen2015resilient} that local sensing is sufficient for each agent to adapt to changes and disruptions in the power grid. It arises from the fact that the measurement of the voltages and power at local buses already contains sufficient information of the measurements from other buses through the physical laws for the learning-based control. Therefore, the multi-agent learning provides a distributed approach for subsystems to coordinate and adapt to each other, which is more resilient to agent's internal faults and malicious attacks.

Looking from the system level, many emerging sophisticated learning paradigms become applicable to develop multi-agent learning-based resilience mechanisms. 
For example, \emph{federated learning} is a distributed machine learning paradigm where all learning agents collaboratively learn a shared prediction model without exchanging the local data. It has been used in the industrial Internet of Things to enable secure and privacy-preserving data access \cite{hou2021mitigating}, energy market prediction for efficient resource allocation \cite{wang2021electricity}, and vehicular networks for intrusion detection \cite{zhang2018distributed}. The application of federated learning in distributed energy systems can coordinate the sub units toward the common objective such as power management and scheduling \cite{wang2020aebis} and energy resource distribution \cite{lee2022federated} without leaking the local data. Therefore, federated learning provides a distributed learning mechanism for task-level coordination as well as safety and privacy guarantees in the learning, which improves the resilience to malicious data access and data breach attacks.

\emph{Meta-learning} provides another efficient way for a learning agent to adapt to different tasks and other agents. Meta-learning first learns a base model to fit all possible tasks, and then only uses a small amount of data to achieve fast adaptation to a specific task. Meta-learning has been used to investigate the energy dispatch mechanism in self-powered and sustainable multi-agent systems \cite{Munir2021multi} and develop power grid emergency control to maintain system reliability and security \cite{huang2022learning}. 
In an ICS, such as distributed energy systems, individual subsystems can have different operational modes. For example, a power generating subsystem can operate at full load, half load, and zero loads. Different modes correspond to different OT dynamics. Therefore, other agents can use meta-learning to first maintain a general model of the power generating subsystem and then fast adapt to the specific operation mode, providing a more efficient learning mechanism. Besides, many subsystems share similar subsystem-level objectives. For example, the wind, nuclear, and coal power generating subsystems all seek to generate power. Therefore, an individual subsystem can leverage meta-learning to form a model for other subsystems and then fast adapt for cooperation. Meta-learning can also be used to estimate the attacker's model for resilience planning. When facing similar cyber attacks, the network defender can learn a general attacker's model using meta-learning and then updates the model with the new network data to develop effective defense strategies.

\emph{Stackelberg learning} is also a useful tool for security and resilience planning when facing external attacks. It involves two players, a leader and a follower, as in Stackelberg games. The leader can be a network defender who anticipates the attacker's (follower) action to make resilient defense strategies \cite{li2018false}. The leader can also be a system operator who maintains the system performance under the attacker's (follower) attack \cite{an2020stackelberg}. The associated Stackelberg game can be also extended to multiple defenders \cite{smith2014multidefender} to model multiple subsystems in ICSs. The learning agents in an ICS act as leaders and learn to protect the subsystems and the ICS from malicious attacks.

However, there are still several challenges with multi-agent learning despite its application for resilient designs. The first challenge is heterogeneity. An ICS is characterized by its heterogeneous subsystems. They need customized learning mechanisms to achieve their distinct objectives. For example, consider the wind power generating subsystem and the power delivery subsystem. Two subsystems have different OT and task objectives. The former focuses on smooth power generation while the latter cares about the power delivery on demand. Abrupt changes in the delivery side can affect the generating subsystem, while an unstable power generating subsystem can also jeopardize the stability of power grids.

The sampling efficiency comes as the next challenge. A learning agent needs sampled network data to make operational decisions and adapt to other subsystems. However, the required samples differ in tasks and subsystems. Even within the same subsystem, the IT and OT system require different sampling strategies. For example, the sampling interval for physical control in the OT system should be finer than the cyber detection in the IT system because a large sampling interval can destabilize the OT system. Time-sensitive subsystems, such as power generating subsystems, require more sampled data than time-insensitive subsystems, such as logistics subsystems. A learning agent must decide on efficient strategies to sample information from itself and other agents. Without adequate samples, the learning may not achieve sufficient resilience to deal with attacks and disturbances. Redundant sampling, however, can lead to latency and waste of memory shortage. Hence learning mechanisms need to consider the trade-off between sampling efficiency and system resilience.

The third challenge is the uncertainties in the learning. A learning agent forms a belief in other agents to anticipate their behaviors in face of uncertainties. An effective learning mechanism should help the agent form a correct belief. Incorrect ones can lead to destabilization and cascading errors in ICSs. Despite many approaches to creating the belief in learning, for example, computing empirical opponent's strategy from historical data or no-regret learning, effective belief-forming strategies should also be task-specific. There are no general rules on belief and anticipation formation. A learning agent needs a context-driven approach to learning the resilient operational strategy.

Although facing challenges, multi-agent learning for ICS resilience is still promising. It allows agents to learn on their own in a hierarchically structured way, and the entire system achieves its goal in the end. Single-point failure in learning will trigger immediate responses from other agents that will respond to the failure to prevent the system-level breakdown.

\section{Resilience of Distributed Energy Systems: A Conceptual Case Study} \label{sec:case}
This section elaborates on the role of learning-based mechanisms in improving the resilience of distributed energy systems in modern power grids. 

\begin{figure}[h]
    \centering
    \includegraphics[scale=0.23]{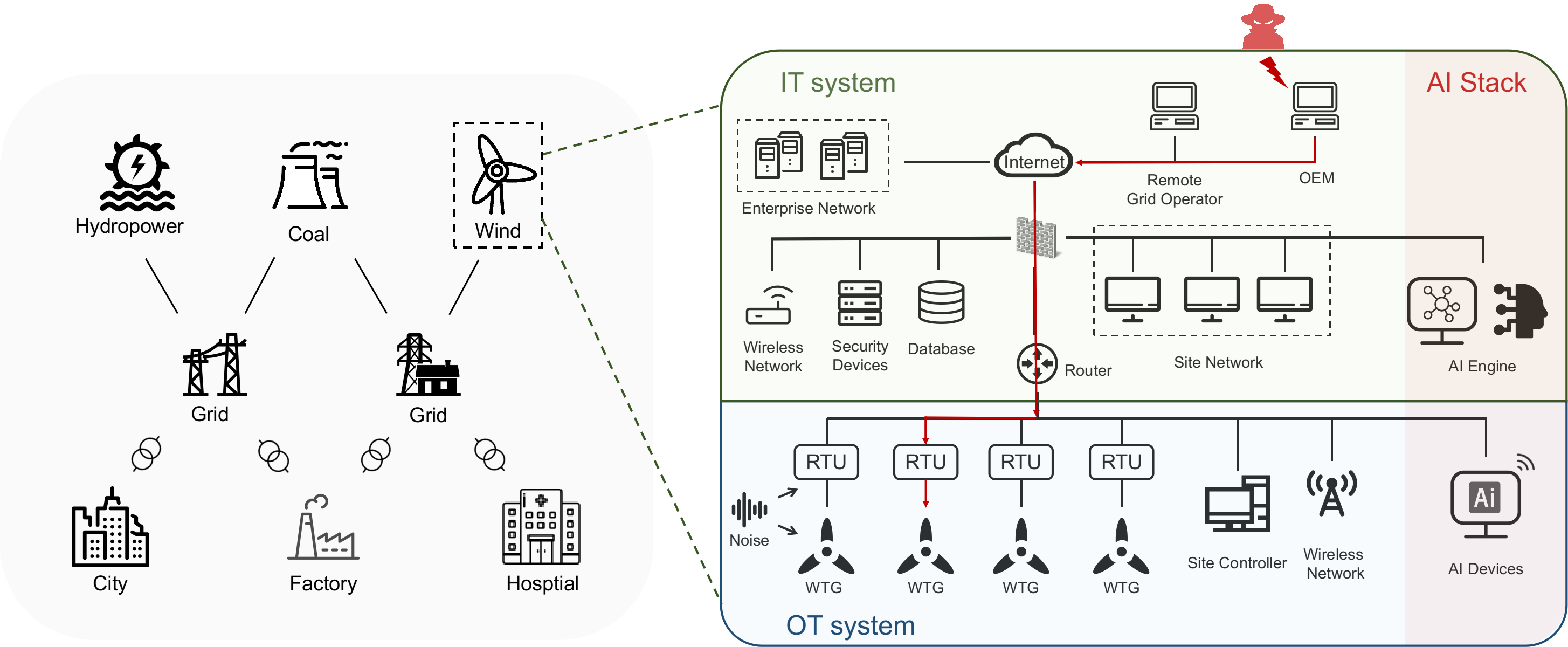}
    \caption{The left part is an example of distributed power systems. A zoom-in illustration of a wind power subsystem is juxtaposed on the right. An attacker can penetrate the IT systems and launch attacks to damage the OT units. Environmental noise can also affect wind turbine performance. The AI stack in the cyber and physical system aims to detect, monitor, and mitigate potential attacks. (OGM: Original Equipment Manufacturer; RTU: Remote Terminal Unit; WTG: Wind Turbine Generator)}
    \label{fig:case_study}
\end{figure}

Distributed energy systems are critical ICSs that generate and deliver power to support indispensable industrial production and transportation activities. They consist of a diverse number of subsystems at the edge, such as solar and wind power systems and local smart grids, working together to achieve a reliable, dependable, and resilient modern power grid. The distributed nature of the grid and the modern integration with IT have brought many concerns about cyber attacks, which can disrupt and degrade the system performance (e.g., the quality and the availability of the power supply). The IT and OT system architecture of a wind turbine is illustrated in Fig.~\ref{fig:case_study}. Attackers can penetrate the IT system and damage the OT units, including sensors, actuators, and controllers. External disturbances, such as noise and device faults, can also hurt the wind turbine performance.

Wind power subsystems generate renewable energy from wind turbines and deliver the power to the grid. The controller center acts as the IT system for power generation scheduling and remote monitoring. At the same time, the OT system is composed of a group of wind turbine generators. Cyber attacks can attack and penetrate the IT system of wind power subsystems and further disrupt sensor signals or disable actuators and controllers to undermine normal turbine operations and power output. Other attacks and failures in the wind power subsystem such as connectivity issues, and turbine damage are also discussed in \cite{barker2020resilient}.

One way to model the confluences between the IT system on the OT is to use a Markov jump linear system (MJLS) that captures the interdependencies between the cyber state, $\theta_t$ and the physical state $x(t)$.    \cite{peng2020distributed,zhu2013resilient,chen2015resilient,zhu2011robust,xu2016cross,chen2017interdependent,chen2022cross}. More formally, we let $x(t)$ be (physical) turbine state, including the rotor speed and output power, and let $\theta_t$ be the operational state, including the normal operation and failure states. The dynamics of the states are given by
\begin{equation}
\label{eq:turbine_dynamics}
\begin{split}
    \dot{x}(t) &= A(\theta_t) x(t) + B(\theta_t) u(t) + E(\theta_t) \hat{v}(t), \\
    y(t) &= C(\theta_t) x(t).
\end{split}
\end{equation}
Here, $u(t)$ and $y(t)$ are control inputs and observations of appropriate dimensions; $A(\theta_t), B(\theta_t), E(\theta_t), C(\theta_t)$ are linearized system matrices of appropriate dimensions at the current wind turbine operational status. The impact of random noise $\hat{v}(t)$ from the environment on the wind turbine dynamics is captured by the matrix $E(\theta_t)$. The dynamics of the operational state $\theta_t$ follow a Markov stochastic process. The transition from one state to the other depends on the behaviors of the cyber system.

Cyber attacks to the IT system can lead to failures in the OT assets, including sensors, actuators, controllers, and communications between them. The attacks can create an effect on the system matrices $C(\theta_t), A(\theta_t), B(\theta_t)$ in the control dynamics \eqref{eq:turbine_dynamics} correspondingly. The attacks and failures discussed in \cite{barker2020resilient} can also be reflected by the change of system matrices. We use the set $\mathcal{S} = \{0,1,\dots, n\}$ to index scenarios of no attack and $n$ attacks. The attacker selects from $\mathcal{S}$ to damage wind power subsystems. Since perfect security is cost-prohibitive, the defender allocates resources in the cyber system to mitigate cyber attacks. Therefore, the attacker and the defender play a non-cooperative security game $\mathbf{G}$.
We denote $\mathbf{g}, \mathbf{f} \in [0,1]^{|\mathcal{S}|}$ as the strategy profile of the attacker and the defender, respectively. Let $g_i, i\in \mathcal{S}$,  denote the probability of the attacker using attack $i$ while $f_j, j\in\mathcal{S},$ denote the probability of the defender prepares for attack $j$. Here, $\mathbf{g}$ and $\mathbf{f}$ reside in an $(n+1)$-dimensional simplex. The defender's utility, denoted by $U_{ij}$ and the attacker's cost, denoted by $C_{ij}$, are defined in \eqref{eq:defender} and \eqref{eq:attacker}, respectively, as follows. 
\begin{equation}
\label{eq:defender}
    U_{ij} := \alpha P_{ij} - C_{d,j},
\end{equation}
\begin{equation}
\label{eq:attacker}
    C_{ij} := \alpha P_{ij} + C_{a,i}.
\end{equation}
Here, $P_{ij}$ is the average output power from wind turbines; $C_{d,j}$ and $C_{a,i}$ are constant defense and attack costs, respectively. The non-cooperative game models the rational attacker and strategic interactions in the cyber defense, and they are solved using bilevel programming. Note that the game can be solved using learning-based approaches such as fictitious play \cite{shen2007adaptive} and no-regret learning \cite{xu2020distributed}. The defender and follower can learn from each other's historical actions and find the equilibrium solution.

The IT and OT systems are closely interdependent. On the one hand, the defense strategy determines the transition rules of the operational state $\theta_t$ in the dynamics \eqref{eq:turbine_dynamics}. On the other hand, the output power of the OT system affects the defense utility in the security game $\mathbf{G}$ \eqref{eq:defender}-\eqref{eq:attacker}. The more power wind turbines produce, the more valuable and critical these wind turbines are, and the more defense reward the defender can attain by protecting them.
The system-of-system equilibrium has been proposed as the holistic solution to cyber attacks \cite{zhu2021control,chen2019games,huangyh2020dynamic}. The equilibrium solution is compatible with both IT and OT systems: it can stabilize wind turbines and maximize the defense reward. An iterative learning algorithm is proposed to find the equilibrium solution, which creates a cyber-physical co-learning paradigm. The IT system first generates a defense strategy and sends it to the OT system. Then, the OT system finds a stabilized controller and informs the IT system of the output power. The interdependent process iterates until a compatible equilibrium solution is reached. 

The system performance is measured by the wind turbine's output power and operation pitch angle to assess the effectiveness of the co-learning framework. The resilience is demonstrated in three aspects. First, the framework enables the recovery of the system performance to the pre-attack level under simulated cyber attacks. Second, the framework allows the system performance to return to the expected value more quickly compared with the case with cyber-unaware defense. Third, the framework shows a significantly improved performance compared with the one with cyber-unaware defense.

Although the previous resilient co-learning framework contains model-based characterizations, many learning-based methods can be easily incorporated into the framework to develop pure learning-based resilient mechanism. For example, in the OT system, the learning-based control methods can be used to stabilize wind turbines and meet load frequency \cite{yan2018data}. Using deep RL and the input-output data, the OT can learn a model that overcomes the nonlinearity of wind turbine dynamics and a stabilized policy that guarantee a stable output of wind turbine. Besides, the learning also reduces the computational burden for the control, especially when the real physical model is complex.
In the IT system, the cyber learning methods discussed in Section \ref{sec:learning} can be readily adopted in the framework. For instance, the IT system can form a belief on the attacker's strategy and estimate the the empirical attack and defense costs $C_{a,i}$ and $C_{d,j}$ using historical attack data or ethical footprinting. It is common that the learning take a feedback structure despite the focus of the applications. Discussed in \cite{huang2022reinforcement},  feedback-enabled learning provides theoretical foundations to develop IT solutions for moving target defense \cite{zhu2013game}, honeypot configurations \cite{huang2019honeypot}, human-machine teaming \cite{huang2022radams}. Equilibrium concepts are useful to predict the consequences of the strategic learning in games. The saddle-point equilibrium between the attacker and the defender indicates the quality of the IT defense mechanism. The iterative learning at the OT and the strategic learning at the IT lead to interdependent co-learning that converges to a system-of-systems equilibrium.

The discussed framework demonstrates the interdependency between the IT and OT systems in a wind power subsystem. It can serve as a building block for creating multi-agent learning-based resilience mechanisms to achieve MAR. The subsystem-level learning-based framework can be readily extended to other subsystems because of its generality, although other subsystems have different security requirements and performance specifications, including the load, the reliability of energy sources, and the network connectivity.

\section{Conclusion} \label{sec:conclusion}
This chapter discusses distributed multi-agent learning mechanisms to achieve resilient design for Industrial Control Systems (ICSs). We first introduce a Multi-Agent System (MAS) framework to provide a two-level perspective of ICSs. At The system level, an ICS is divided into functional subsystems with different task objectives. Inside each subsystem, we have Information Technology (IT) and Operation Technology (OT) systems operating together to accomplish the specific task. Next, the introduction of an AI stack enables the data analytics and computational intelligence for subsystems in ICSs to detect, respond, and recover from malicious attacks and disturbances. With the assistance of the AI stack, subsystems interact and coordinate with each other to provide distributed intelligence and resilience. Within each subsystem, the cyber-physical co-learning provides effective detection and mitigation strategies to defend against cyber attacks and stabilize physical plants. The learning at the two levels constitute distributed multi-agent learning mechanisms to achieve the holistic ICS resilience.
We provide an overview of related learning-based methods for resilient ICS design, including supervised and unsupervised learning and game-theoretic learning. We also use a case study in distributed renewable energy systems to elaborate the multi-agent learning mechanism and its effectiveness in providing ICS resilience.

Multi-agent learning for control system resilience is still in its infancy. Despite challenges in developing effective and resilient learning mechanisms as discussed in Section \ref{sec:learning}, multi-agent learning is promising to enable distributed resilience for large-scale and heterogeneous ICSs. It automates the agents to learn on their own in a hierarchically structured way to achieve the goal of the entire system. Many emerging learning paradigms including federated learning, meta-learning, and Stackelberg learning are also applicable to achieve  multi-agent ICS resilience.

\bibliographystyle{IEEEtran}
\bibliography{mybib}
\end{document}